\documentclass[%
 aip,groupedaddress,
 jmp,%
 amsmath,amssymb,
preprint,%
]{revtex4-1}

\usepackage{graphicx}% Include figure files
\usepackage{dcolumn}% Align table columns on decimal point
\usepackage{bm}% bold math
\def\grl{{\em Geophys. Res. Lett.}}
\def\jgr{{\em J. Geophys. Res.}}
\def\prl{{\em Phys. Rev. Lett.}}

\def\sci{{\em Science}}
\def\pop{{\em Phys. Plasma}}
\def\pof{{\em Phys. Fluid}}

\def\spose#1{\hbox to 0pt{#1\hss}}
\def\approxlt{\mathrel{\spose{\lower 3pt\hbox{$\sim$}}
        \raise 2.0pt\hbox{$<$}}}
\def\approxgt{\mathrel{\spose{\lower 3pt\hbox{$\sim$}}
        \raise 2.0pt\hbox{$>$}}}

\def\multleft#1{\hbox to size{\vbox {\halign {\lft{##}\cr #1}}\hfill}\par}
\def\multright#1{\hbox to size{\vbox {\halign {\rt{##}\cr #1}}\hfill}\par}

\def\boxit#1{\vbox{\hrule\hbox{\vrule\kern3pt\vbox{\kern3pt
          #1 \kern3pt}\kern3pt\vrule}\hrule}}

\begin{document}

\preprint{AIP/123-QED}

\title{Buneman instability in a magnetized  current-carrying plasma with velocity shear}% Force line breaks with \\
%\thanks{Footnote to title of article.}

\author{H. Che, M. V. Goldman, D. L. Newman}
%\author{B. Author}%
% \email{Second.Author@institution.edu.}
\affiliation{ Center of integrated Plasma Studies, University of Colorado, Boulder, CO, 80309-0390, USA}
%Authors' institution and/or address%\\This line break forced with \textbackslash\textbackslash
% \homepage{http://www.Second.institution.edu/~Charlie.Author.}
%\affiliation{%
%Second institution and/or address%\\This line break forced% with \\
%}%

\date{\today}% It is always \today, today,

\begin{abstract}
Buneman instability is often driven in magnetic
reconnection. Understanding how velocity shear in the
beams driving the Buneman instability affects the growth
and saturation of waves is relevant to turbulence,
heating, and diffusion in magnetic reconnection. Using a
Mathieu-equation analysis for weak cosine velocity shear
together with Vlasov simulations, the effects of shear on
the kinetic Buneman instability are studied in a plasma
consisting of strongly magnetized electrons and cold
unmagnetized ions. In the linearly unstable phase, shear
enhances the coupling between oblique waves and the
sheared electron beam, resulting in a wider range of
unstable eigenmodes with common lower growth rates.  The
wave couplings generate new features of the electric
fields in space, which can persist into the nonlinear
phase when electron holes form. Lower hybrid
instabilities simultaneously occur at
$k_{\shortparallel}/k_{\perp} \sim \sqrt{m_e/m_i}$ with a
much lower growth rate, and are not affected by the
velocity shear.
\end{abstract}

\maketitle

\section{Introduction}
\label{intro}
The Buneman instability has been extensively studied in
theory and simulation since it was discovered in 1958
\cite{buneman58prl}. It is well-known in one-dimensional
(1D) theory that once the relative drift of ions and
electrons exceeds the threshold of approximately twice
the electron thermal velocity, the interactions between
waves and electrons will lead to the growth of the
Buneman instability\cite{davidson70prl,papa77rgsp}. Lower hybrid
instabilities lie on an oblique branch of current-driven
instabilities\cite{miki74book,kindel71jgr} with
$k_{\shortparallel}/k \sim \sqrt{m_e/m_i}$. Their
relation to Buneman instabilities has not been clarified
yet in two-dimensional (2D) spectral space. Recently,
interest in Buneman and lower hybrid instabilities has
been renewed because of their importance in magnetic
reconnection\cite{drake03sci,jovanovi05pop,shokri05pop,fuji06pop,dyrud06jgr,mcmillan07pop,goldman08grl,che09prl,che10grl,yoon10pop}.
Magnetic reconnection is one of the most relevant
mechanisms associated with explosive events in nature and
in laboratory experiments, such as solar flares,
substorms in the magnetosphere, and sawtooth crashes in
fusion experiments. 

\begin{figure}
\includegraphics[scale=0.45,trim=30 310 0 50,clip]{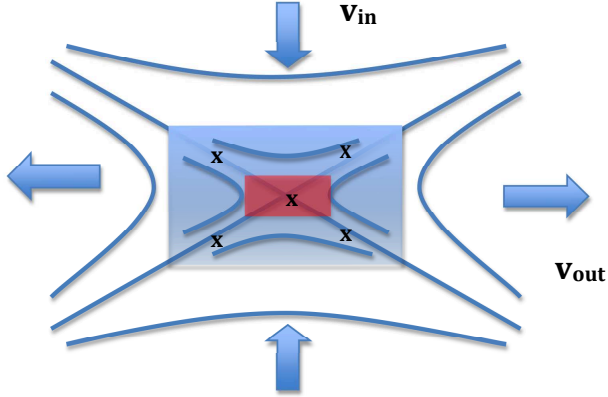}
\caption{Illustration of magnetic reconnection. The red
  color indicates the electron diffusion region around
  the x-line with a scale of the electron skin depth
  $c/\omega_{pe}$. The semi-opaque blue color indicates the
  ion diffusion region with a scale of the ion skin depth
  $c/\omega_{pi}$. The inflow velocity is $v_i$ and the
  outflow velocity is $v_o$. The stars ``*'' mark some of the possible regions
  where electron holes can develop. }
\label{illu}
\end{figure}

 Magnetic reconnection can convert
magnetic field energy into thermal and kinetic
energy.  Oppositely directed magnetic fields merge and
lead to the release of a significant fraction of the
stored magnetic energy, which produces two regions of
fast outflow (see Fig.~\ref{illu}). One of the most
important problem in understanding magnetic reconnection
is determining what mechanisms can facilitate
reconnection fast enough to explain the explosive events
observed. From magnetohydrodynamic (MHD) models, the
convection and diffusion of the magnetic field can be
described in terms of the magnetic Reynolds number $R_m
=4\pi L_0 V_0/c^2 \eta^2 $, where $L_0$ and $V_0$ are
respectively the typical plasma velocity and magnetic
field spatial length. If $R_m \ll 1$, the collisional
resistivity can effectively dissipate the magnetic
field's energy and facilitate fast reconnection. However,
the following question remains: What drives fast
reconnection if the collision-induced resistivity is not
sufficiently large (i.e. $R_m \gtrsim 1$)? This is a
condition common in both space and laboratory
plasmas. One of the most promising and physically
interesting mechanism proposed to answer this question is
turbulence-induced dissipation facilitating fast
reconnection\cite{sagdeev62book,davidson75pof,papa76pop,huba77grl,kulsrud98pop}.

For turbulence-induced dissipation to work, the
turbulence has to be generated spontaneously during the
magnetic reconnection.  In the diffusion region of
magnetic reconnection, current sheets form as a result of
changes in the magnetic field configuration. The intense
thin current sheets which develop in strong guide field
reconnection can drive streaming instabilities
\cite{drake03sci,mcmillan07pop,che09prl,che10grl} and
electron velocity shear
instabilities\cite{ganguli88pof,prichett93pof,drake94prl,nish03pop,ferraro04pop,che10nat}
etc.  The latest 3D PIC simulations show that anomalous
momentum transport generated by an electromagnetic
electron velocity shear instability can influence the
magnetic reconnection process\cite{che10nat}, but if
turbulence resistivity can affect reconnection rates
is still unclear. A deeper understanding of current-driven
electrostatic instabilities is required. Buneman and
lower hybrid instabilities are two of the most common
\textit{electrostatic} instabilities driven within
current sheets in magnetic reconnection. They can produce
electron holes and strong turbulent heating near the
x-line and near the separatrix
\cite{farrell02grl,matsumoto03grl,ferraro04pop,vai04grl,cattell05jgr,khot10prl,che10grl}
(Fig.\ref{illu}).

Current sheets become thinner and thinner during
reconnection, and velocity shear is generated in the
current sheets regardless of the initial velocity
distribution. Extensive studies on new instabilities
driven by electron velocity shear within current sheets
have been performed\cite{ganguli88pof,prichett93pof,drake94prl,ferraro04pop,eliasson06pop}.
However, only a few studies have been carried out
regarding how velocity shear affects the classical
Buneman and lower hybrid instabilities\cite{reitzel98pop}
although their role is important to the understanding of
turbulence-induced dissipation processes in reconnection.
 
 In this paper we investigate the role of weak velocity
 shear on Buneman and lower hybrid instabilities in
 2D. Specifically, we study a plasma model consisting of
 strongly magnetized electrons and cold unmagnetized
 ions. A Mathieu equation analysis, first proposed by
 Goldman\cite{goldman04math,goldman05ttsp}, is used and
 the results are compared with those of Vlasov
 simulations\cite{newman07pop,newman08pop}.The results
 we obtain from the Mathieu equation analysis are found
 to be consistent with the results from the Vlasov
 simulations with weak initial velocity shear, validating
 the approximations we adopted in the analytical method.
 We compare the Buneman instability driven by electrons
 with a uniform velocity drift (i.e., the classical
 Buneman instability) with the instability driven by
 drifting electrons with a weak cosine velocity shear. We
 find that the Buneman instability is no longer a purely
 magnetic field-aligned instability as commonly assumed
 in uniform beam\cite{buneman58prl,yoon08pop}. The
 velocity shear enhances the interplay between oblique
 waves and electrons and produces a wide range of
 eigenmodes with a common growth rate lower than in the
 uniform-drift case.  As shear increases, the fastest
 growing mode changes from a parallel plane wave to an
 eigenmode with significant oblique wavenumber content,
 as discussed in Appendix \ref{map}. The shear does not
 significantly change the growth rate of the co-existing
 lower hybrid instability, which is much weaker than that
 of the Buneman instability under our assumptions. We
 obtain the eigenfunctions in 2D spectral space from the
 Mathieu equation analysis. These eigenfunctions show
 that the shear greatly modifies the 2D spatial
 structures of the instability-induced electric fields.

\section{Basic equations and solutions }
\subsection{Basic equations}
\label{mathieu} 
We assume 1) the electrons are strongly magnetized such
that $r_e = v_{te}/\Omega_{e} \ll \lambda_e =
v_{te}/\omega_{pe} $, where $r_e$ is the electron
cyclotron radius, $\lambda_e$ is the Debye length,
$\omega_{pe}$ is electron plasma frequency,
$v_{te}=\sqrt{\textit{k}T/m_e}$ is the electron thermal
velocity, and $\textit{k}$ is Boltzmann's constant. 2)
the kinetic electrons are constrained to move along
magnetic field lines while the ions are treated as
unmagnetized and cold (i.e., the ratio of ion and
electron temperature satisfies $T_i\ll T_e$). The
validity of this assumption requires that the wave phase
speeds are larger than ion thermal velocity.

We study the instabilities in the 2D $z$--$y$ plane. The
electrons move along the $z$ direction with drift
velocity $v_b(y)$. The magnetic field is treated as
infinite in the $z$ direction for electrons and zero for
ions, with no initial electric field ($E_0=0$). The
initial ion drift velocity is also zero.  The electron
drift velocity is modulated by a weak cosine velocity
shear:
\begin{equation}
v_b(y)=v_0 [1+ \varepsilon cos2\pi(y/L_y - 0.5)], y \in
[0, L_y],
\label{shear}
\end{equation}
 where $\varepsilon$ is a small quantity and $L_y$ is the
 (spatial) periodicity of the shear profile.

Because the velocity shear is a function of $y$, the
perturbed electric fields generated by instabilities are
also functions of $y$. With the assumptions of cold ions
and infinitely magnetized electrons, the perturbed and
unperturbed functions do not depend on $v_y$.  We perturb
the Vlasov equation with $f_q=f_{0q} +\delta
f_q=f_{0q}(v_{qz},y)+\delta f_q(v_{qz},y)e^{i(k_z z -
  \omega t)}$, where $q$ represents electrons ($e$) or
ions ($i$). The electrostatic potential is $\delta
\Phi=\phi(y)e^{i(k_z z - \omega t)}$. The cold
($|\omega/k|>v_{ti}$; here $v_{ti}$ is the ion thermal
velocity) initial ion distribution function can be
written as $f_{0i} =n_0\delta(v_ {iy})\delta(v_{iz}) $.
To properly build our model, the initial electron
distribution function is approximated by a 1D drifting
kappa-function\cite{summers91pof} with $\kappa=1$:
\begin{equation}
  f_{0e}=\dfrac{2n_0}{\pi v_{te}} \left[1+\dfrac{(v_{ez} -
        v_b(y))^2}{v_{te}^2}\right]^{-2}.
\end{equation}

The kappa-functions used to represent the electron
distribution are quasi-Maxwellian at sub-thermal
velocities, but with power-law rather than exponentially
decreasing tails at supra-thermal velocities. Such
power-law tails are a common feature of measured
distributions in collisionless space plasmas. In our
model the kappa distribution also simplifies the
functional form of the susceptibilities, and thus enables
us to obtain a specific mathematical differential
equation with general solutions. We have compared our
theoretical predictions to the results of 2D Vlasov
simulations using both Maxwellian and kappa electron
distributions, and find that the behavior in the two
simulations is qualitatively equivalent. Only results
from the simulations with Maxwellian electron
distributions are presented in this paper.

 After inserting the perturbations and initial conditions into
 the first order linear Vlasov equation and Poisson
 equation,
 \begin{align}
\frac{\partial\delta f_q }{\partial t} &+ v_{qz}
\frac{\partial\delta f_q}{\partial z} - \frac{q}{m}
\frac{\partial \delta \phi}{\partial z} \frac{\partial
  f_{0q}}{ \partial v_{qz}} = 0,\\ &- \triangle
\delta\Phi (y, k_z,\omega) = 4 \pi (\rho_e + \rho_i),
\end{align}
we obtain
\begin{eqnarray}
\frac{\partial^2 \phi}{\partial y^2} - k_z^2(1 +
\frac{\chi_{ezz}}{1+\chi_{izz}})\phi = 0,\\ 
\chi_{ezz} =-
\dfrac{\omega_{pe}^2}{n_0 k_z^2}\int_C
\dfrac{\textsl{f}_{0e}(v_{ez})
  dv_{ez}}{(\frac{\omega}{k_z}- v_{ez})^2},
\label{chi_e}\\ 
\chi_{izz}
= -\dfrac{\omega_{pi}^2}{n_0 k_z^2}\int_C
\dfrac{\textsl{f}_{0i}(v_{iz})
  dv_{iz}}{(\frac{\omega}{k_z} - v_{iz})^2},
\label{pos1}
\end{eqnarray}
 where $C$ indicates integration along the Landau contour
 in the complex plane, and $\rho_e$ and $\rho_i$ are the
 electron and ion charge density respectively.
 
Upon substituting $f_{0e}$ and $f_{0i}$ into
(\ref{chi_e}) and (\ref{pos1}) we find
 \begin{eqnarray}
\chi_{izz} = -\frac{\omega_{pi}^2}{\omega^2},\\
\chi_{ezz} = - \frac{\omega_{pe}^2}{k_z^2 v_{tez}^2} \frac{\xi + 3i}{(\xi +i)^3},
\end{eqnarray}
where $\xi = [\omega - k_z v_b(y)]/k_z v_{tez}$.
 
 We normalize the quantities by defining $\omega
 \rightarrow \hat{\omega} \equiv \omega/\omega_{pe}$;
 $k_z \rightarrow K_z \equiv k_z
   v_0/\omega_{pe}$; $v_{tez} \rightarrow u_e \equiv
 v_{tez}/v_0$; $S \equiv (L_y \omega_{pe}/\pi
   v_0)^2$, so that
\begin{eqnarray}
\frac{\partial^2 \phi}{\partial \theta^2} - SK_z^2(1 +
\frac{\chi_{ezz}}{1+\chi_{izz}})\phi = 0,\\ \chi_{izz} =-
\hat{\omega}_{pi}^2/\hat{\omega}^2,\\ \chi_{ezz}=-\frac{1}{K_z^2
  u_e^2} \frac{\xi + 3i}{(\xi + i)^3},
\end{eqnarray}
where $\theta \equiv \pi(y/L_y - 0.5), \in
[-\pi/2,\pi/2]$. For later use, we also define the
normalization $k_y\rightarrow K_y\equiv k_yv_0/\omega_{pe}$.

Finally, we expand $\chi_{ezz}$ in $\varepsilon$ to the first-order
and obtain
 \begin{equation}
\frac{\partial^2 \phi}{\partial \theta^2} + (a - 2 q\cos2\theta)\phi=0,
\label{ma1}
\end{equation}
where $\phi$ is a function of $\theta$, the parameters
$a$ and $q$ are complex, and defined as
\begin{equation}
\eta=\frac{\hat{\omega}-K_z}{K_z u_e},  \qquad a =
-SK_z^2(1+\frac{\chi_{ezz0}}{1+\chi_{izz}})
\label{params1}
\end{equation}
and
\begin{equation}
q=-\frac{\varepsilon S}{u_e^3} \frac{\eta +4i}{(\eta
    +i)^4(1+\chi_{izz})},
\label{params2}
\end{equation}
so that
\begin{equation}
\chi_{ezz0}=\chi_{ezz}\vert_{\varepsilon=0}=-\frac{1}{K_z^2
  u_e^2} \frac{\eta + 3i}{(\eta +i)^3}.\\
\label{xez0}
\end{equation}

Equation~(\ref{ma1}) is the well-known Mathieu equation.

 \subsection{Solutions of the Mathieu Equation}
 \label{solmathieu}
 The Mathieu equation (\ref{ma1}) is written in standard
 form\cite{bogo61book}. The quantity $a$ is the
 characteristic value (or eigenvalue) and the parameter
 $q$ is defined in (\ref{params2}).  For specific pairs
 $(a,q)$, the Mathieu equation has a unique analytical
 solution (eigenfunction) $\phi$, which can be written as
 $\phi_r(\theta) = e^{ir\theta} \tilde\phi_{r}(\theta)$,
 where $r$ is an integer or a rational number. The value
 $r$ is called the characteristic exponent, and
 $\tilde\phi_{r}$ is a complex function of $\theta$ that
 can be either even or odd. For periodic boundary
 conditions, $\phi_r$ has period $\pi$, and r is required
 to be integer. In the case of the electric field, $r$ is
 more restrictively required to be an \textit{even}
 integer.  If $q =0$, Eq.~(\ref{ma1}) reduces to the
 standard oscillator equation, where $r=\sqrt{a}$ and the
 solutions reduce to $\phi_r \propto
 \cos(\sqrt{a}\theta)$ (even function) and
 $\sin(\sqrt{a}\theta)$ (odd function).
 
 The fact that $\phi_r =
 e^{ir\theta}\tilde\phi_{r}(\theta)$ suggests that the
 parameter $r$ is related to to the perpendicular
 wavenumber $k_y$. For the case $q=0$, a map $r=r(k_y)$
 can be established. In equation (\ref{ma1}), $r$ is
 related to $\theta\equiv\pi(y/L_y -0.5)$ so that
 $\theta$ scales with the width of the box $L_y$; thus
 $k_y=r\pi /L_y $. However, for the case $q\neq 0$, the
 correspondence between $r$ and $k_y$ is nontrivial and
 needs an alternative treatment (see Appendix~\ref{map}).
 
 We solve equation (\ref{ma1}) with the given periodic
 boundary conditions in $r$--$K_z$ space for
 $\varepsilon=0$ and 0.2. The case of $\varepsilon =0$
 implies $q=0$, which corresponds to uniform velocity
 drift. For specificity, we choose the following
 parameter values: $v_0 = 5 v_{te}$, $L_y =256
 \lambda_e$, and the ratio of ion to electron mass is
 1836, so that $S=266$. These parameters are the same as
 those used in the Vlasov simulations described in the
 next section. To allow comparison between the analytical
 solutions and our Vlasov simulations, we use the
 electric field eigenmodes $E_z^{eig}$ and $E_y^{eig}$
 instead of $\Phi$:
 \begin{align}
\delta\Phi_r &=\phi_r(\theta)e^{i(k_z z -\omega t)},
\\ E_z^{eig} &= -Re\left(\frac{\partial \delta
  \Phi_r}{\partial z}\right) =
-Re\left(ik_z\phi_r(\theta)e^{i(k_z z -\omega t)}\right),
\\ E_\theta^{eig} &= -Re\left(\frac{\partial \delta
  \Phi_r}{\partial \theta}
\right)=-Re\left(\frac{\partial \phi_r}{\partial
  \theta}e^{i(k_z z -\omega t)}\right), \\ E_y^{eig} &=
\frac{\pi}{L_y} E_{\theta}^{eig}.
\label{mae2d}
\end{align}

 \begin{figure}
\includegraphics[scale=0.5,trim=20 0 0 0,clip]{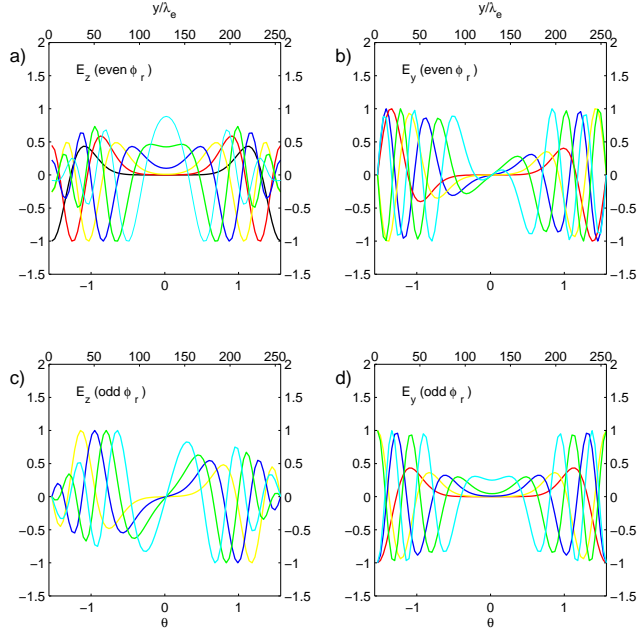}
\caption{Eigenmodes of $E_z$ and $E_y$ for $z=0$ and
  $t=0$ calculated from both even and odd eigenfunctions
  of the electrostatic potential $\phi_r$ with even
  integer $r$ obtained from Mathieu equation
  (\ref{ma1}). The lines colored black, red, yellow,
  green, blue, and cyan represent $r=$0, 2, 4, 6, 8, and
  10, respectively. For odd eigenmodes, $r=2,4,6,8,10$. }
\label{eigfun}
\end{figure}

\begin{figure*}
\includegraphics[scale=0.7,trim=80 10 0 0,clip]{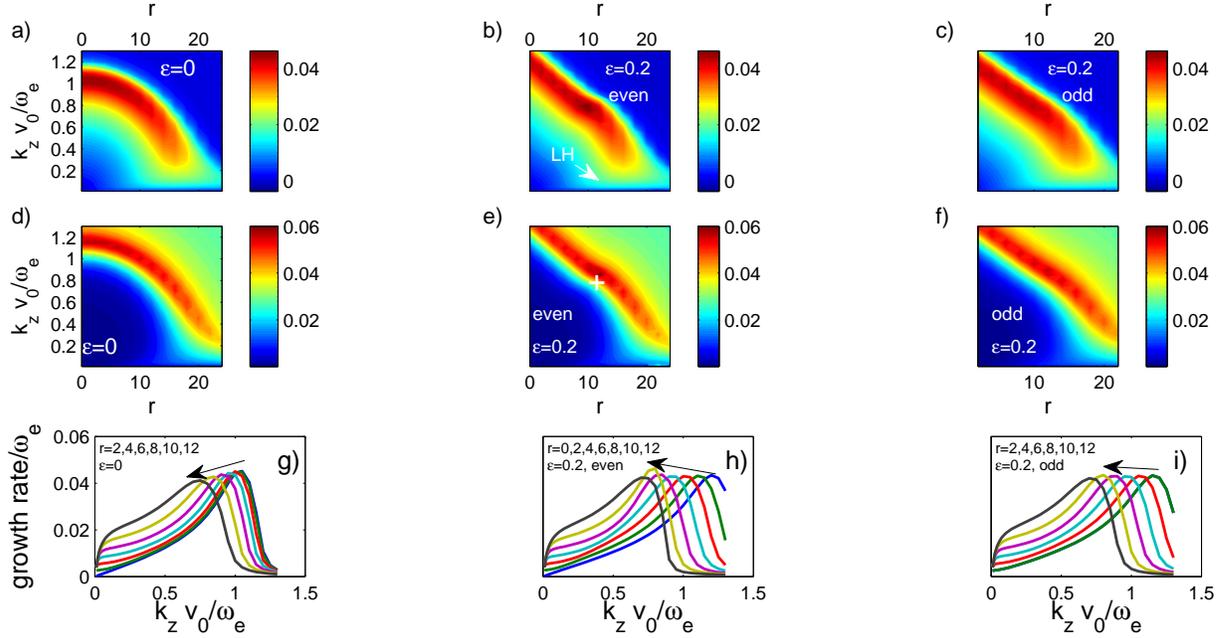}
\caption{Theoretical growth rate in $r$--$K_z$ space in
  panel (a) and frequency in panel (d) for case
  $\varepsilon =0$; growth rate in panel (b) and
  frequency in panel (e) for even $\phi_r$ for
  $\varepsilon=0.2$ while growth rate in panel (c) and
  frequency in (f) are for odd $\phi_r$ for
  $\varepsilon=0.2$.  respectively. The peak of the
  growth rate in (a) is around $(r, K_z) \sim (0, 1)$,
  the peak of the growth rate in (b) is around $(r, K_z)
  \sim (10,0.8)$ and the peak of the growth rate in (c)
  is around $(r, K_z)\sim (2, 1.2)$. The symbol ``+" in
  panels (e) indicate the position of the peak growth
  rate for $\varepsilon=0.2$ in $r$--$K_z$ space. Panels
  (g), (h) and (i) show 1D growth rates vs. $K_z$ for
  cases $\varepsilon =0$ and $\varepsilon=0.2$. The
  arrows in (g), (h) and (i) indicate the directions in
  which $r$ increases.  }
\label{macspec}
\end{figure*}

For periodic boundary conditions we require not only that
the eigenfunction $\phi$ be periodic, but also require
that $E_y^{eig}$ be periodic so that the electric fields
are continuous at the boundary. As we have mentioned only
the eigenfunctions with even integer r can satisfy these
continuity requirements.  For odd eigenfunctions $\phi_r$
we require $r\neq 0$.  For each value of $r$, the
parallel wavenumber $K_z = k_z v_0/\omega_{pe}$ is chosen
to maximize the eigenmode growth rate for that $r$. (See
Fig.~\ref{macspec}(b,c) below).

Figure \ref{eigfun} shows $E_y^{eig}$ and $E_z^{eig}$ at
$z=0$ and $t=0$ for $\varepsilon = 0.2$.  For $E_y^{eig}$
and $E_z^{eig}$ obtained from even $\phi_r$, $r$=0, 2, 4,
6, 8, and 10; and from odd $\phi_r$, $r=2,4,6,8,10$. When
$z=0$ and $t=0$, $E_y^{eig} \propto -\partial
\phi_r/\partial y $ and $E_z^{eig} \propto \phi_r$. We
scale $E_z^{eig}$ and $E_y^{eig}$ so that their maximum
amplitudes are normalized to unity based on the
assumption that all of the electric-field perturbations
with different wavelength initially have approximately
the same amplitude. For even $\phi_r$, we can see that
$E_y^{eig}$ vanishes at $\theta=0$. By contrast, for even
$\phi_r$, $E_z^{eig}$ vanishes at $\theta=0$ for small
values of $r=0, 2, 4$ (black, red, and yellow lines,
respectively) while $E_z^{eig}$ peaks for the larger
values of $r$=8 and 10 (green and cyan lines,
respectively). Comparing electric fields from odd
$\phi_r$, $E_y^{eig}$ is similar to $E_z^{eig}$ from even
$\phi_r$, but the peak value at $\theta=0$ of $E_y^{eig}$
(r=10) is much lower.  $E_z^{eig}$ obtained from odd
$\phi_r$ vanishes at $\theta =0$ for all $r$. These
properties can produce specific features in the 2D
electric fields that distinguish the electric fields in
the sheared case from those produced by the instability
driven by uniform electron drift.  For $\varepsilon=0$,
the initial electric-field eigenmodes at $z=0$ are
proportional to either $\cos(r\theta)$ or $\sin(r\theta)$
with even integer $r$.

We obtained the linear frequency and growth rate of the
unstable eigenmodes for both even and odd $\phi_r$ for
each pair $(r,K_z)$ for both $\varepsilon=0$ and
$\varepsilon=0.2$.  We show the results in
Fig.~\ref{macspec}.  Panels (a), (b) and (c) show the
growth rate (imaginary part of the frequencies) in
$r$--$K_z$ space, and panels (d), (e) and (f) show the
corresponding real wave frequencies. Panels (g), (h) and
(i) are 1D growth rates vs. $K_z$ for some specific $r$
values for cases $\varepsilon=0$ and $\varepsilon=0.2$,
respectively. The arrows in (g), (h) and (i) indicate the
direction of increasing $r$.

The unstable eigenmodes for the uniform beam in panel (a)
($\varepsilon=0$) are concentrated in two distinct areas,
representing two different instabilities, one strong and
the other weak. The stronger instability is the parallel
mode at $K_y \sim 0, K_z\sim 1$ with peak growth rate
$\sim0.05 \omega_{pe}$, comparable to the cold plasma
limit of the growth rate for the fast-growing mode of the
Buneman instability $ \gamma_{max} \sim
\frac{\sqrt{3}\omega_{pe}}{2}(\frac{m_e}{2m_i})^{1/3}$,
where $\omega_{pe}$ is the electron plasma frequency. The
growth rate peaks around $K_z \sim 1$, the same as in the
uniform cold plasma beam limit where $v_0 =
k_z/\omega_{pe}$, and the frequency is $\omega \sim 0.03
\omega_{pe}$ [panel (d)], which is close to the
cold-plasma limit $\omega_r \sim
\frac{\omega_{pe}}{2}(\frac{m_e}{2m_i})^{1/3}$. The peak
growth rate of the weaker lower-hybrid instability is
located near $r=14$ and $K_z=0.03$, with a growth rate of
$\sim 0.01$. Since $r=14$ corresponds to $K_y = k_y
v_0/\omega_{pe} \sim 1$, we have $k_z/\sqrt{k_z^2+k_y^2}
\sim k_z/k_y \sim 0.03 \sim \sqrt{m_e/m_i}$, which
corresponds to the ratio of the parallel to the
perpendicular component of the wave vector of the lower
hybrid instability. The corresponding frequency is $\sim
0.01 \omega_{pe} \sim \omega_{pi}$ [panel (d)], also
around the cold plasma limit for the lower hybrid
instability
$\omega_{lh}=\omega_{pi}/(1+\omega_{pe}^2/\Omega_{e}^2)^{1/2}$,
where $\omega_{pi}$ is the ion plasma frequency and
$\Omega_{e}$ is the electron cyclotron frequency
(effectively infinite in our model). The growth rate of
the Buneman instability decreases steeply with $K_z$ when it
passes the peak, but in panels (g), (h) and (i), we see
that the 1D growth rate curves for $r=6, 8,10,12$ exhibit
a plateau at very small $K_z <0.1$ in all three cases;
these plateaus are caused by the presence of the lower
hybrid instability.

Lower hybrid waves are oblique and can interact with both
ions and electrons. For the ordering $\Omega_i \ll
\omega_{LH} \ll \Omega_e$, an approximation in which the
electrons are treated as magnetized and the ions as
unmagnetized is often justified. Under this approximation
electrons resonate with waves through their motion
parallel to the magnetic field, satisfying
$\omega/k_\parallel\sim v_d$.  At the same time, ions
resonate with waves through their motion both parallel
and perpendicular to the magnetic field satisfying
$\omega/k\sim v_i$. The waves resonating with drifting
electrons (in the ion rest frame) and ions can lead to
the growth of both Buneman and lower hybrid
instabilities, with the former dominated by ion motion
\textit{parallel} to $\mathbf{B}$ and the latter by ion
motion \textit{perpendicular} to $\mathbf{B}$. Oblique
lower hybrid waves therefore transfer the parallel
momentum of electrons predominantly to the perpendicular
momentum of ions.  Thus the lower hybrid instability can
be thought as the oblique limit of the Buneman
instability. In our model, electrons are assumed to be
infinitely magnetized while ions are assumed to be
unmagnetized and cold, which is the extreme limit of this
model. The transfer of momentum does not increase the
temperature of ions in our approximation.

When comparing Fig.~\ref{macspec}(b,e) and (c,f) for the
sheared beam $\varepsilon = 0.2$, to
Fig.~\ref{macspec}(a,d) for uniform beam, it is obvious
that the velocity shear has changed the location of the
most unstable eigenmodes for even potential
eigenfunctions $\phi_r$ [panel (b)] to $(r, K_z)
\approx(10, 0.8)$, with a growth rate close to $0.05
\omega_{pe}$. A number of eigenmodes with $r<14$ have a
comparable growth rate. The corresponding frequency shown
in panel (e) is $\sim 0.04 \omega_{pe}$ (with the
fastest-growing mode indicated by the white cross), which
is slightly higher than without shear.  The lower hybrid
instability, however, is not affected by the velocity
shear, which might be due to our assumption of cold ions.
The unstable eigenmodes for odd potential eigenfunctions
$\phi_r$ in panel (c) with $r<14$ have nearly equal
growth-rate maxima, with that of the $r=2$ mode being
slightly higher than for other unstable modes, but lower
than the growth rate of the fastest-growing ($r=10$)
eigenmode for even $\phi_r$. Thus we can conclude that
the most unstable mode is still associated with the
Buneman instability. The velocity shear enhances the
coupling between oblique waves and electrons.  The
parallel phase speeds for both cases are around $0.03
v_0$, matching the cold plasma limit.

\section{Results of Vlasov simulations and Comparisons with theory}
We have carried out two 2D Vlasov simulations with two
different velocity shears: $\varepsilon =0 $ and
$\varepsilon =0.2$. The simulation box size is $L_y
\times L_z = 256\lambda_e \times 800 \lambda_e$. The
velocity range is $[-15, 15] v_{te}$. The grids in phase
space are $(L_z, v_{ez}, v_{iz}, L_y, v_{ey},v_{iy} )=
(512, 128, 128, 64, 1,64)$.  The time step is about
$0.626 \omega_{pe}^{-1}$ and The total simulation time is
$\omega_{pe} t \sim 1004$. We initialize the simulations
with Maxwellian distribution functions for both electrons
and ions, and work in the initial ion rest frame. The
electrons are given a mean drift $v_0 = 5 v_{te}$ with a
weak cosine velocity shear (for $\varepsilon\neq 0$) as
shown in Eq.~(\ref{shear}). The ion-electron mass ratio
is $1836$. We assume that the magnetic field along $z$ is
infinite for electrons, so the electron distribution
function is restricted to a three-dimensional
($z$--$y$--$v_z$) phase space. The ions are assumed to be
unmagnetized. The electron-ion temperature ratio is $5$
so that the ions are relatively cold.

\begin{figure}
\includegraphics[scale=0.6,trim=100 20 0 100,clip]{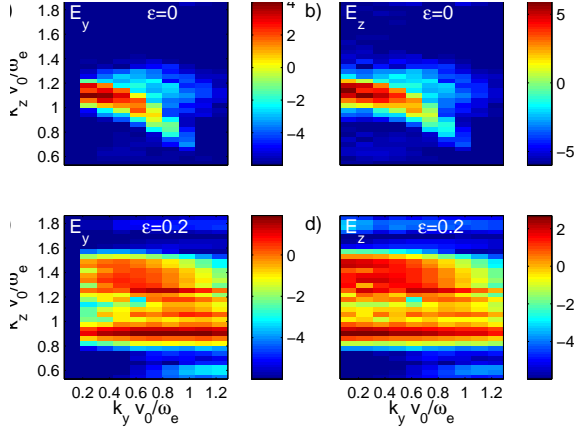}
\caption{Panels (a) and (b) are, respectively, power
  spectra of $E_y$ and $E_z$ in logarithmic scale in 2D $k$ space at
  $\omega_{pe} t=552$ for the $\varepsilon =0$ Vlasov
  simulation; Panels (c) and (d) are the same as (a) and
  (b) for $\varepsilon=0.2$. [Note: some labels are cut
    off in the embedded PDF version of this figure.]}
\label{simspec}
\end{figure}

In \S\ref{solmathieu} we have shown that both Buneman and
lower hybrid instabilities can co-exist at the same time,
but the lower hybrid instability is much
weaker. Therefore in the Vlasov simulations we do not
expect to observe the lower hybrid instability during the
linear growth phase. In the case of $\varepsilon=0.2$,
the fastest-growing unstable Buneman mode is the $r=10$
potential eigenmode for even $\phi_r$. This mode is
dominant in the simulation while other unstable
eigenmodes for $r=0,2,4,6,8$ for even $\phi_r$ compete
with the unstable eigenmodes for odd $\phi_r$. To verify
this we show the power spectra of the electric fields
$|E_y(\mathbf{k})|^2$ and $|E_z(\mathbf{k})|^2$ in
logarithmic scale at $\omega_{pe} t=552$ in both
simulations in Fig.~\ref{simspec}.

 In the case of $\varepsilon=0$ [Fig.~\ref{simspec} (a,
   b)], the dominant $k$-space mode is parallel to $z$
 (i.e., $k_y\approx0$) with $K_z \sim 1.1$. A Fourier
 transform from the time domain over an interval
 containing $\omega_{pe} t=552$ reveals the frequency to
 be about $0.03\omega_{pe}$. Both the wavevector and
 frequency are consistent with the theoretical
 predictions for the Buneman instability with uniform
 electron velocity drift. The 2D power spectra
 $|E_y(\mathbf{k})|^2$ and $|E_z(\mathbf{k})|^2$ for
 $\varepsilon=0.2$ are shown in panels (c) and (d). We
 see that the strongest unstable modes for $E_z$ are now
 near $K_z = 0.9$, which is consistent with the
 fastest-growing unstable $r=10$ eigenmode for even
 $\phi_r$ at $K_z\sim 0.8$ in Fig.~\ref{macspec}b, and
 span a large range in $K_y$. The Fourier transform from
 the time domain shows the frequency increases slightly
 to $0.04\omega_{pe}$.  In this case $r$ has a more
 indirect relation to $K_y$. Here we will use the mapping
 of eigenmodes from $r$ to $K_y$ discussed in Appendix
 \ref{map}.  For the globally fastest-growing eigenmode
 for even eigenfunctions $\phi_r$ at $r=10$ and
 $K_z=0.8$, the spectrum of $E_z(K_y,K_z)$ peaks near
 $K_y=0.2$.  The neighboring $E_z$ even eigenmodes at
 $r=$4, 6, 8, and 12, which have almost as high a growth
 rate, exhibit multiple maxima in their $K_y$ spectra,
 with peaks both at low $K_y$ $(<0.2)$ and high $K_y$
 $(>0.5)$ This behavior can be compared with that of
 corresponding \textit{synthetic} $K_z$--$K_y$ spectra in
 Fig.\ref{rkymap1} and \ref{rkymap} of Appendix
 A. It is therefore reasonable to conjecture that the
 strongest modes of $E_z$ in Fig.~\ref{simspec} are
 produced by the higher-$r$ eigenmodes for even $\phi_r$
 (e.g., $r=8$--12) for which $E_z^{eig}$ peaks near the
 center in Fig.~\ref{eigfun}a, although the width of the
 $E_z$ spectrum in $K_z$ from Fig.~\ref{simspec}d is
 slightly narrower than would be expected based on the
 locus of growth-rate maxima (as a function of $r$) in
 Fig.~\ref{macspec}b. The modes in Fig.~\ref{simspec}
 with $K_z \gtrsim 1.1$, by contrast, come from the
 low-$r$ eigenmodes for even $\phi_r$ (e.g., $r=0$--4),
 for which $E_z^{eig}$ peaks near the edge in
 Fig.~\ref{eigfun}a. The growth rate of unstable
 eigenmodes for odd $\phi_r$ compete with the lower $r$
 eigenmodes for even $\phi_r$ and $K_z\sim 0.8$--$1.4
 $. These modes spread the spectrum of $E_z$.  A similar
 argument can be used to explain the power spectrum of
 $E_y$. The $E_z$ and $E_y$ spectra for $\varepsilon=0.2$
 peak at two distinct ranges of $K_z$, which come,
 respectively, from the two regions where $\partial
 v_b/\partial y$ vanishes: the center and edge of the
 box.  At these locations $v_b$ is respectively 20\%
 higher and 20\% lower than for
 $\varepsilon=0$. Therefore, the dominant $K_z$ in each
 region should be $\sim$20\% lower and $\sim$20\% higher
 than for $\varepsilon=0$ if we apply the cold plasma
 relation $k_z= \omega_r/v_{b}$.

\begin{figure}
\includegraphics[scale=0.5,trim=0 0 0 0,clip]{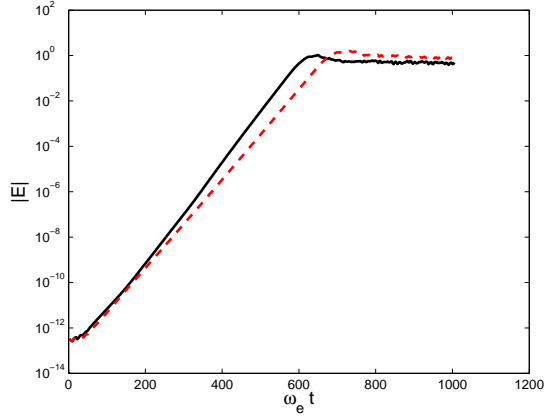}
\caption{The growth in time of the spatially averaged
  electric field $\langle |E_z| \rangle$ in Vlasov
  simulations for $\varepsilon=$0 (black solid curve) and
  $\varepsilon=0.2$ (dashed red curve). }
\label{et}
\end{figure}

The time evolution of the spatially averaged electric
fields in both simulations shown in Fig.~\ref{et} can be
divided into two stages: linear and nonlinear.  In the
nonlinear stage the electric field reaches its peak and
saturates via electron trapping. In the case of
$\varepsilon=0.2$, electric fields become nonlinear
slightly later than when $\varepsilon=0$ due to the small
difference between the growth rates of the Buneman
instability in these two cases.  An effective growth rate
$\gamma$ can be determined from the simulation fields
during the linear phase through the relation $\gamma
\Delta t \equiv \ln (E/E_0)$.  For $E\sim 1$ and $E_0
\sim 10^{-13}$, $\omega_{pe} \Delta t=$600 and 700
corresponding to $\gamma \approx$ 0.05 and 0.043 for
$\varepsilon=$0 and 0.2, respectively. These values are
consistent with the results from linear kinetic theory
shown in Fig.~\ref{macspec}.

\begin{figure}
\includegraphics[scale=0.7,trim=30 0 0 70,clip]{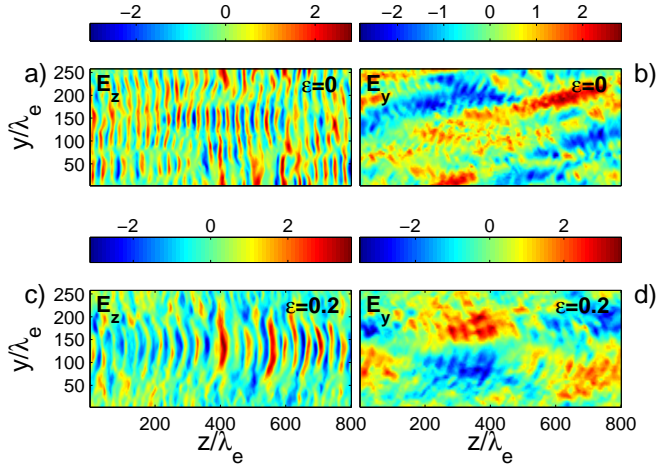}
\caption{Panels (a) and (b) are the Vlasov-simulation
  electric fields $E_z$ and $E_y$ for $\varepsilon =0$ at
  $\omega_{pe} t=679$. Panels (c) and (d) are the same as
  (a) and (b) for $\epsilon =0.2$ and $\omega_{pe} t
  =753$. The two times correspond to approximately the
  same late nonlinear stage. }
\label{ne2d}
\end{figure}

 In Fig.~\ref{ne2d}, we show the
electric fields $E_z$ and $E_y$ from Vlasov simulations
in $z$--$y$ space. Panels (a) and (b) are for
$\varepsilon=0$ at $\omega_{pe} t = 679$ and panels (c) and
(d) are for $\varepsilon=0.2$ at $\omega_{pe} t=753$, when
the simulations for both cases are in a similar (late)
nonlinear stage. These figures show localized and intense
structures, indicative of electron trapping. 

\begin{figure}
\includegraphics[scale=0.45,trim=30 50 0 20,clip]{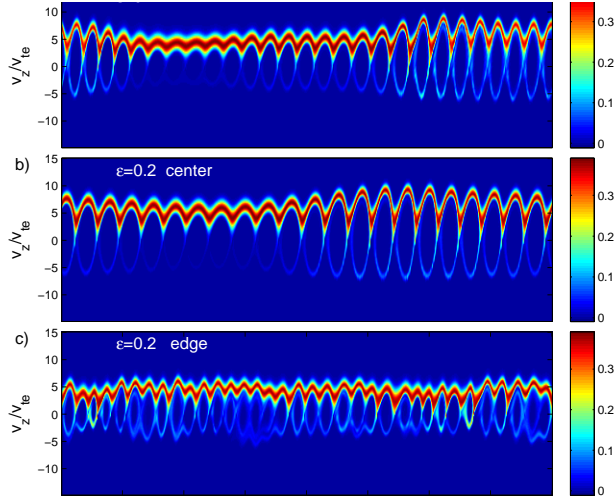}
\caption{Panels (a), (b) and (c) are the
  Vlasov-simulation electron distribution functions in
  $z$--$v_z$ phase space. (a) is at $y=128\lambda_e$ for
  $\varepsilon=0$ at $\omega_{pe} t=$627; (b) is at the
  center and $y=128\lambda_e$ and (c) is at the edge of
  the $y$ for $\varepsilon=0.2$ at time $\omega_{pe}
  t=$691, respectively. The electron holes indicate that
  both simulations are at approximately the same early
  nonlinear stage.}
\label{holes}
\end{figure}

A comparison at a slightly earlier time of structures in
$z$--$v_z$ phase space is shown in
Fig.~\ref{holes}. Panel~(a) is the electron distribution
function $y=128\lambda_e$ for $\varepsilon=0$ at
$\omega_{pe} t=627$, (b) is at $y=128\lambda_e$ and (c)
is at the edge of $y$ for $\varepsilon=0.2$ at
$\omega_{pe} t=691$.  Again, because the instability with
$\varepsilon=0$ grows faster than with $\varepsilon=0.2$,
the evolution of electric fields and electron holes shown
in Fig.~\ref{holes} are both at approximately the same
(early) nonlinear stage. That the width of electron holes
in (c) is smaller than the width of electron holes in (b)
is consistent with the wavelength of the Buneman
instability in sheared beam. It's interesting that the
electron holes at the edge seems less regular than the
electron holes at the center due to the fact that the
electric field $E_z$ is weaker than the electric field
$E_z$ at the center of $y$.

\begin{figure}
\includegraphics[scale=0.7,trim=60 30 0 80,clip]{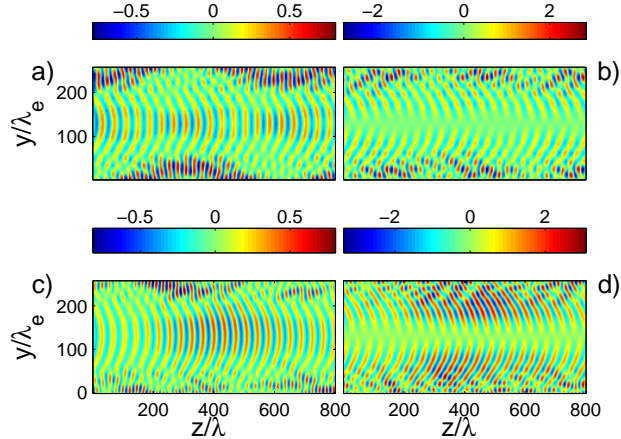}
\caption{Panels (a) and (b) show $E_z$ and $E_y$,
  respectively, in $z$--$y$ space as synthesized from a
  superposition of theoretically determined eigenmodes
  for $\varepsilon=0.2$; Panels (c) and (d) show $E_z$
  and $E_y$, respectively, from the $\varepsilon=0.2$
  Vlasov simulation at $\omega_{pe} t=552$, which is during
  the linear stage.}
\label{e2d}
\end{figure}

  To further demonstrate the eigenmode structure of the
  Buneman instability when $\varepsilon=0.2$, we compare
  2D electric fields $E_y$, and $E_z$ based on the
  Mathieu-equation analysis to the electric fields in the
  linear stage of the corresponding Vlasov simulation at
  $\omega_{pe} t=552$. To do this, we construct a
  wavepacket of eigenfunctions with similar growth rates
  as a superposition of eigenfunctions with different
  weight and phase for $r=0$, 2, 4, 6, 8, 10, and 12 for
  even potential eigenfunctions $\phi_r$ and
  $r=2,4,6,8,10,12$ for odd potential eigenfunctions
  $\phi_r$.  From the theoretical growth rate, we see
  that the features of the electric fields come mainly
  from the eigenmodes at $r=10$ for even potential
  eigenfunction $\phi_r$ and $r=2$ for odd potential
  eigenfunction $\phi_r$.  The $r=10$ even eigenmode is
  dominated by small $K_y$ (long-wavelength) behavior
  near the center ($y=128\lambda_e$) while the $r=2$ odd
  eigenmode is dominated by large $K_y$
  (short-wavelength) behavior near the edges. We draw the
  2D $E_z$ structure synthesized from an eigenmode
  superposition in Fig.~\ref{e2d}(a) and the
  corresponding $E_y$ in (b). Not surprisingly, we see
  that $E_z$ peaks at the center ($y/\lambda_e \sim 128$)
  where the drift is a maximum, while $E_y$ vanishes
  there.  Compared to the electric fields at $\omega_{pe}
  t=552$ in the linear stage of the Vlasov simulation
  shown in Fig. \ref{e2d} (c, d), the superposition of
  theoretical electric fields reproduces the main
  features of the simulation electric fields, but not
  necessarily all of the subtle details, which are
  controlled by the exact amplitudes and phases of the
  different contributing eigenmodes. We show the power
  spectra of the theoretical electric fields in
  \ref{micspec} in Appendix. These features of the
  electric fields persist into the nonlinear stage, even
  after electron holes form (e.g., at $\omega_{pe} t\sim
  700$, when the simulation has evolved into the
  nonlinear stage).  Even though the electric fields
  shown in Fig.~\ref{ne2d}(c, d) become localized, we
  still see the $y$-dependence discussed above.  On the
  other hand, $E_z$ in panel (a) and $E_y$ in panel (b)
  do not show any preferred value of $y$, as expected for
  a uniform (i.e., unsheared) electron beam.
 
\section{conclusion}
In this paper we explored the impact of velocity shear on
the Buneman and lower hybrid instabilities.  We have
studied the unstable modes for electron drifts modulated
by a weak cosine velocity shear. In this case the
dispersion relation can be approximated by the well-known
Mathieu equation, where we used a kappa-function for the
initial distribution of infinitely magnetized electrons
(with cold unmagnetized ions), and we use a Taylor
expansion for the velocity shear.  These approximations
have been validated by comparing the analytical
results with the results of Vlasov simulations. 

The effect of velocity shear is expected to strengthen
the interactions between oblique waves and particles
because shear causes the fastest-growing Buneman mode to
change from a parallel plane wave to an eigenmode
containing oblique as well as parallel Fourier modes. The
shear also distributes the unstable Buneman eigenmodes
over a wider range of eigenvalues, with nearly the same
slightly lower growth rate than in the shear-free
case. We conclude that the growth rate, wavelength, and
orientation of the fastest growing Buneman mode are
controlled by the presence (and amplitude) of shear in
the electron velocity drift. Therefore, velocity shear
can lead to different growth rates and wavelengths of
unstable modes in the direction of the velocity
gradient. The Vlasov simulation show that the electron
holes form at both the center and the edge of the
simulation box. At the center, electron trapping exhibits
a longer wavelength consistent with the fact that the
beam is at its fastest; at the edge, electron trapping
exhibits a shorter wavelength, consistent with the
correspondingly slower beam.

The velocity shear does not affect the growth rate of the
weak lower hybrid instability, in which the interactions
between lower hybrid waves and ions can transfer electron
parallel momentum to the perpendicular motion of the
ions. This momentum transfer can occur without
necessarily affecting the temperature of the ions.

The advantage of our method is that it provides
eigenfunctions for every unstable mode so that we can
investigate both the spectra and 2D spatial structure
of the electric fields. These results can help us to
interpret the physical content of simulations,
experiments and satellite data. Our method has assumed
that the shear is weak.  However, stronger velocity shear
may reveal new physical regimes, including
\textit{shear-driven} as well as \textit{shear-modified}
instabilities.  To address this issue, we are undertaking
a study of how far the Mathieu-equation analysis can be
extended.

\begin{acknowledgments}
This work was supported by NASA MMS-IDS Grant NNX08AO84G. 
\end{acknowledgments}

\appendix
\section{Mapping Mathieu equation eigenfunctions into 
Fourier space }
\label{map}

\begin{figure}
\includegraphics[scale=0.7, trim=50 20 0 80,clip]{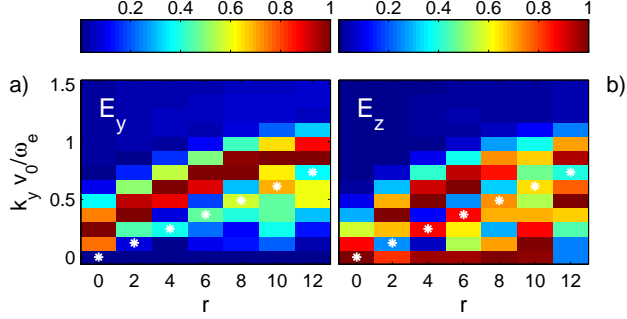} 
\caption{The correspondence between $K_y$ and even integer $r$. Panel
  (a) shows $E_y$ and panel (b) shows $E_z$ for
  $\varepsilon=0.2$. The white stars mark the linear
  correspondence $K_y=r \pi /L_y$ for
  $\varepsilon=0$. }
\label{rkymap1}
\end{figure}

\begin{figure}
\includegraphics[scale=0.5, trim=10 20 0 20,clip]{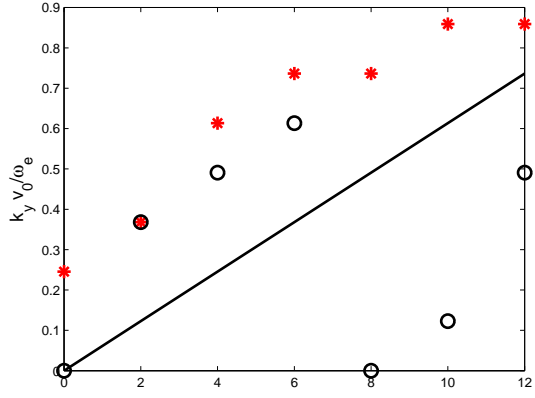} 
\caption{The correspondence between $K_y$ and $r$. The
  black circles are obtained for $E_z$ and red stars are
  for $E_y$ for the case $\varepsilon=0.2$. The black
  line is for $\varepsilon=0$.}
\label{rkymap}
\end{figure}

We build a map between $r$ and $k_y$  by projecting the
\textit{complex} eigenmodes of $E_y(y,k_z)$ and
$E_z(y,k_z)$ at t=0 into the $z$--$y$ plane,
\begin{widetext}
\begin{eqnarray}
E_z^{eig}& =&- Re(ik_z(\cos(k_z z) + i \sin (k_z z))
(Re(\phi_r) + i Im (\phi_r)))\nonumber \\ &=& k_z\cos(k_z
z) Im(\phi_r)+\sin(k_z z) Re(\phi_r),\\ E_y^{eig} &=&-
\frac{\pi}{L_y}Re(ik_z(\cos(k_z z) + i \sin (k_z z))
\frac{\partial}{\partial \theta}(Re(\phi_r) + i Im
(\phi_r)))\nonumber \\ &=&\frac{\pi}{L_y}(k_z\cos(k_z z)
Im(\partial \phi_r/\partial \theta)+\sin(k_z z)
Re(\partial \phi_r/\partial \theta)).
\end{eqnarray} 
\end{widetext}
We then perform a 2D Fourier transformation of
$E_z^{eig}$ and $E_y^{eig}$ in $(K_y, K_z) =
v_0/\omega_{pe} (k_y, k_z)$ space. The results of Fourier
transformation should be the same for even and odd
eigenfunctions. In this appendix, we use even
eigenfunctions $\phi_r$ to analyze the relation between
$r$ and $K_y$. The results show that for
$\varepsilon=0.2$ the corresponding power spectrum is no
longer concentrated on one specific $K_y$ but is instead
spread over a range of $K_y$, as shown in
Fig.~\ref{rkymap1}.
 We take the $K_y$ corresponding to the maximum power
 intensity. On the other hand the correspondence between
 $r$ and $K_y$ is different for $E_y$ and
 $E_z$. Fig~\ref{rkymap} shows this map between $K_y$ and
 $r$. Black circle is for $E_z$. The $E_z$ eigenmodes
 $r=8,10,12$ have $K_z \sim 0.8$ while $r=10$ corresponds
 to $K_y \sim 0.2$, $r=6, 8, 12$ correspond to $K_y \sim
 0.6, 0, 0.5$ respectively.  $r=0$ eigenmode corresponds
 to $K_y \sim 0.1, K_z \sim 1.1$ and others in between
 $K_z \sim 0.8 - 1.1$ and $K_y \sim 0.1-0.6$. The $E_y$
 eigenmodes for $r=0,2,4,6,8,10$ have a relatively linear
 correlation with $K_y$, but for $r=12$, $K_y$ decreases
 from $0.9$ to $0.5$.

\begin{figure}
\includegraphics[scale=0.6, trim=120 150 0 60,clip]{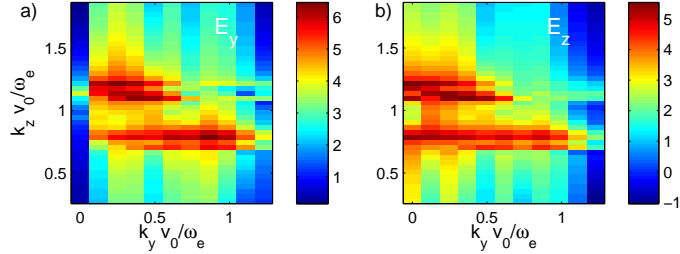} 
\caption{The power spectra for theoretical electric
  fields $E_y$ and $E_z$ shown in Fig.~\ref{e2d} (a,
  b). Panel (a) is for $E_y$ and panel (b) is for $E_z$.}
\label{micspec}
\end{figure}

To provide a fuller picture of the spectra from theory, we
show in Fig.~\ref{micspec} the power spectra of the
superposition of theoretical eigenmode electric fields
$E_y$ and $E_z$ previously shown in Fig.~\ref{e2d}.
We reiterate that the 2D theoretical $E_y$ and $E_z$ only
reproduce the main features of the electric fields and
thus the their power spectra shown in Fig.~\ref{micspec}
can only approximate the realistic power spectra from the
simulation as shown in Fig.~\ref{simspec}. However,
Figs.~\ref{simspec} and \ref{micspec} do both show
the same two distinct ranges of $K_z$.  Again, the
differences are controlled by the exact amplitudes and
phases of the contributing eigenmodes, which we can not
determine theoretically.
%\end{widetext}
%\nocite{*}
%\bibliographystyle{apsrev4-1}
%\bibliography{plasma}% Produces the bibliography via BibTeX.
%

\end{document}